\documentclass[aps,prl,twocolumn,showpacs,floatfix,nobibnotes,superscriptaddress,longbibliography,nofootinbib]{revtex4-1}

\usepackage{inputenc}
\usepackage{amssymb}
\usepackage{graphicx}
\usepackage{verbatim}
\usepackage{epsfig}
\usepackage{bm}
\usepackage{color}
\usepackage{xcolor}
\usepackage{float}
\usepackage{dcolumn}
\usepackage{multirow} 
\usepackage{physics}
\usepackage{braket}

\usepackage[unicode=true,
  linktocpage,
  linkbordercolor={0.5 0.5 1},
  citebordercolor={0.5 1 0.5},
  colorlinks=true,
  linkcolor=blue,
  citecolor=blue,
  urlcolor=blue]{hyperref}

\usepackage{footnote}
\hypersetup{colorlinks=true,
    linkcolor=magenta,
    citecolor=blue,
    urlcolor=cyan,
    pdfpagemode=FullScreen,}

\newcommand{\ra}[1]{\renewcommand{\arraystretch}{#1}}

\usepackage{ulem}
\newcommand\redsout{\bgroup\markoverwith{\textcolor{red}{\rule[0.5ex]{2pt}{1.4pt}}}\ULon}

\renewcommand{\emph}[1]{\textit{#1}}

\begin{document}


\title{Revisiting the helium isotope-shift puzzle with \\improved uncertainties from nuclear structure corrections}


\author{Simone Salvatore Li Muli}
\email{simone.limuli@chalmers.se}
\affiliation{Institut f\"ur Kernphysik and PRISMA$^+$ Cluster of Excellence, Johannes Gutenberg-Universit\"at, 55128 Mainz, Germany}
\affiliation{Department of Physics, Chalmers University of Technology, SE-412 96 G\"oteborg, Sweden}
 
\author{Thomas R.~Richardson}
\email{richardt@uni-mainz.de}
\affiliation{Institut f\"ur Kernphysik and PRISMA$^+$ Cluster of Excellence, Johannes Gutenberg-Universit\"at, 55128 Mainz, Germany}

\author{Sonia Bacca}
\email{s.bacca@uni-mainz.de}
\affiliation{Institut f\"ur Kernphysik and PRISMA$^+$ Cluster of Excellence, Johannes Gutenberg-Universit\"at, 55128 Mainz, Germany}
\affiliation{Helmholtz-Institut Mainz, Johannes Gutenberg-Universit\"at Mainz, D-55099 Mainz, Germany}

\begin{abstract}
Measurements of the difference between the squared charge radii of the helion ($^3$He nucleus) and the $\alpha$-particle ($^4$He nucleus) have 
 been characterized by longstanding tensions,
recently spotlighted in the 3.6~$\sigma$ discrepancy of the  extractions from ordinary atoms versus those from muonic atoms~\cite{Sch23}.
Here, we present a novel analysis of uncertainties in nuclear structure corrections that must be supplied by theory to enable the extraction of the difference in radii from spectroscopic experiments.
We use modern Bayesian inference techniques to quantify uncertainties stemming from the truncation of the chiral effective field theory expansion of the nuclear force for both muonic and ordinary atoms.
With the new nuclear structure input, the helium isotope-shift puzzle cannot be explained, rather it is reinforced to a 4~$\sigma$ discrepancy.
\end{abstract}

\maketitle

\textit{Introduction---}
\label{introduction}
Historically, atomic physics has played a central role in shaping modern physics.
Explaining the  gross features of the hydrogen-atom spectrum led to the development of quantum mechanics \cite{Sakurai,Griffiths},
studying its fine-structure details inspired  relativistic quantum mechanics~\cite{Dir28}, and the discovery of the Lamb shift \cite{Lam47}  gave rise to the theory of quantum electrodynamics (QED)~\cite{Tom48,Sch48,Fey49}.
While the hydrogen atom is still a protagonist~\cite{Bey17,Fle18,Bez19}, other simple atomic systems, such as  hydrogen-like ions~\cite{Kra19} or two-electron atoms~\cite{Pac17,Pat21}, have entered the scenery of high precision
laser spectroscopy. 
Today, atomic physics is experiencing an exciting time, when not only  fundamental constants---such as the Rydberg constant~\cite{Tie21}---are determined with better than ever precision, but  the comparison of results from multiple experiments on a variety of simple and calculable  systems allows for fruitful intersections with particle, hadronic, and nuclear physics~\cite{Ant22,Antognini:2022jec}. 
In fact, high precision laser spectroscopy can be used  as a rigorous test of the Standard Model and has the potential to constrain sources of beyond the Standard Model physics. 
Furthermore, muonic atom spectroscopy allows for precise determinations of the size of the nucleus~\cite{Poh10,Ant13,poh19, Kra21,Sch23} and nuclear polarizability effect~\cite{poh19} because the mass of the muon, approximately 200 times that of the electron, makes the muon very sensitive to nuclear structure.

Recently, the difference between the squared charge radius of the helion ($^3$He nucleus) and the $\alpha$-particle ($^4$He nucleus), defined as $\delta r^2= r^2_\text{ch}(^3{\rm He}) -r^2_\text{ch}(^4{\rm He})$, has attracted significant attention. 
This difference can be extracted from the isotope shifts of the  $2^3{\rm S}\rightarrow 2 ^1 {\rm S}$ transition~\cite{Roo11,Ren18,Van23}, the $2^3{\rm S}\rightarrow 2 ^3 {\rm P}$ transition~\cite{Shi95,Can12,zhe17}, and the $2^1{\rm S}\rightarrow 2 ^1{\rm D}$   transition~\cite{Hua20}, but the obtained values vary quite substantially, see Fig.~5 in Ref.~\cite{Sch23} for a summary. 
In particular, results from the same group in Refs.~\cite{Roo11,Ren18,Van23} are in disagreement; however, a re-analysis confirmed that differences are understood and that the most recent experiment~\cite{Van23} supersedes the older two \cite{Roo11,Ren18}.
The final value inferred from Ref.~\cite{Van23} is $\delta r^2=1.0757(15)$ fm$^2$.

The difference in the charge radii obtained from the isotope shift in ordinary atoms can be compared to the results obtained from muonic atoms. 
In the latter, the absolute values of the individual radii can be extracted from Lamb-shift measurements.
The sizes of $^3$He~\cite{Kra21} and $^4$He~\cite{Sch23} were recently measured by the CREMA collaboration resulting in $\delta r^2= 1.0636(6)^{\rm exp}(30)^{\rm theo}$ fm$^2$ \cite{Sch23}.
Comparing this value to the most recent measurement based on the isotope shifts in ordinary atoms, in particular the measurement of Ref.~\cite{Van23}, reveals a 3.6~$\sigma$ discrepancy.

The error bars in radii extractions from muonic atom experiments are largely dominated by uncertainties coming from theoretical calculations of nuclear structure corrections, 
which have previously been evaluated by performing few-body calculations with two different parameterizations of the nuclear force.
In this work, we provide an update of the nuclear structure corrections based on chiral effective field theory ($\chi$EFT) in conjunction with modern Bayesian uncertainty quantification techniques.


\textit{Nuclear structure corrections}---
The observed energy spectrum of an atom differs from that obtained via the solution of the Schrödinger or Dirac equation in a static Coulomb potential because of quantum electrodynamics (QED), nuclear recoil, and nuclear structure effects.
The difference of two energy levels in an atom containing a ${}^A$He nucleus is parameterized as
    \begin{equation}
    \Delta \text E = \delta_{{\rm QED}} + \mathcal{C}  r_\text{ch}^2 +  \delta_{\text{NS}}\, .
    \label{EQ: 1}
    \end{equation}
In Eq.~\eqref{EQ: 1}, $\delta_{\text{QED}}$ comprises purely QED and nuclear recoil effects, and the second term is a nuclear finite-size effect where $\mathcal{C}$ is a known constant for each of the transitions considered. 
The last term, $\delta_{\text{NS}}$, contains nuclear structure corrections that begin with the exchange of two photons.
At the level of two-photon exchange, these corrections enter through the nuclear matrix elements of the forward virtual Compton tensor \cite{Friar_PRC_13} and constitute the dominant source of uncertainty in the extraction of $r_{\text{ch}}$.

The $\delta_{\text{NS}}$ is expanded as
    \begin{align}
        \label{EQ:ns}
        \delta_{\text{NS}} & = \delta_{\text{TPE}} + \delta_{\text{3PE}} + \delta_{\text{EVP}} + \delta_{\text{MSEVP}} + \cdots
    \end{align}
where the terms include  two-photon exchange $\delta_{\text{TPE}}$, three-photon exchange $\delta_{\text{3PE}}$,  electron vacuum polarization $\delta_{\text{EVP}}$, and muon self energy and vacuum polarization  $\delta_{\text{MSEVP}}$ corrections. For muonic atoms, the last three terms have been recently discussed in Ref.~\cite{Pac23}\footnote{Note that in \cite{Pac23}, the last two terms are labelled sightly differently, with eVP and $\mu$SE, respectively.}.
In this work, we focus on the two-photon exchange term, which can be written as \cite{Ji_13,Ji_18}
    \begin{align}
        \delta_{\text{TPE}} & = \delta^A_{\text{TPE}}+ \delta^N_{\text{TPE}} \nonumber \\
        &=\delta^A_{\text{Zem}} + \delta^A_{\text{pol}} + \delta^N_{\text{Zem}} + \delta^N_{\text{pol}} \, ,
        \label{Eq:tpe_sep}
    \end{align}
where $\delta^A_{\text{Zem}}$ ($\delta^N_{\text{Zem}}$) denotes nuclear (single-nucleon) elastic or Zemach contributions, and $\delta^A_{\text{pol}}$ ($\delta^N_{\text{pol}}$) denotes nuclear (single-nucleon) inelastic or polarizability contributions.
In both muonic and ordinary atoms, the Zemach contribution will be canceled by a piece of the polarizability correction.
In the remainder of this work, we denote the two-photon-exchange contributions in muonic (ordinary) atoms by $\delta^{A}_{\text{TPE}, \mu}$ ($\delta^{A}_{\text{TPE}, e}$).

In muonic atoms, the nuclear excitation energy is generally much smaller than the muon mass. In the so-called $\eta$-expansion formulation \cite{Pachucki_PRA_15,Ji_13,Ji_18,Friar_PRC_13,LiMuli22},
the two-photon-exchange correction  can be decomposed in multipoles leading to a dipole 
contribution and higher-order terms.
Explicit expressions for these terms as well as the Zemach term may be found in  Ref.~\cite{Ji_18} and are therefore not repeated here. 

For electronic atoms, the opposite scenario is realized where the nuclear excitation energy is generally much larger than the electron mass.
In this case, the leading contribution from nuclear structure effects is \cite{Friar_PRC_97, Pac07} 
    \begin{align}
        \delta^{A}_{\text{TPE}, \, e} = -\frac{2}{3} m (Z\alpha)^2 \phi_{nS}^2 \ \tilde\alpha_{\text{pol,e}} 
        \label{tpe_e}
    \end{align}
with
\begin{equation}
     \tilde \alpha_{\text{pol,e}}= \sum_{N \neq 0} \lvert \bra{N} \mathbf{D} \ket{0} \rvert^2 \Biggl[ \frac{19}{6\omega_N} + \frac{5 \ln(2\omega_N/m)}{\omega_N} \Biggr] \ ,
     \label{tilde_pol}
\end{equation} 
where $\mathbf{D}$ is the electric-dipole operator. 
We neglect higher order terms  as they are found to be small ($1\%$ in the deuterium $1$S-$2$S transition~\cite{Friar_PRC_97}), but will later include an uncertainty to account for this assumption.

We evaluate $\delta^{A}_{\text{TPE}, \, \mu}$ and $\tilde \alpha_{\text{pol,e}}$ by solving the few-nucleon Schrödinger equation to obtain $\ket{0}$ and $\ket{N}$ via the effective interaction hyperspherical harmonics (EIHH) method, which is very accurate for three- and four-body problems \cite{EIHH, Kam01, LiMuli21}. 
We use nuclear forces derived from $\chi$EFT, which is a generalization of chiral perturbation theory, an EFT based on the spontaneously broken approximate chiral symmetry of quantum chromodynamics, to multi-nucleon systems~\cite{wei91,wei92,mac11,Epe22}. 
There are many families of $\chi$EFT interactions available in the literature.  Here,  for the first three orders we use the local formulation of Refs.~\cite{gez14,lyn16,lyn17} (with the cutoff  $r_0=1.0$~fm),   which we recently implemented in the EIHH code~\cite{LiMuli21,Ach23}, 
 while for the highest order we take the non-local two-body force from~Ref.~\cite{Ent03} supplemented by the local three-body forces from Ref.~\cite{Nav07}.


\textit{Bayesian uncertainty quantification}---
The truncation error of the $\chi$EFT expansion is quantified by calculating nuclear structure corrections at different orders. Bayesian inference can be used to analyze the convergence pattern of any observable 
with respect to a reference ${\mathcal O}^{\text{Ref}}$
\begin{align}
{\mathcal O}={\mathcal O}^{\text{Ref}}~\sum_{n=0}^{\infty} c_n Q^n \ , \label{EQ: 12}
\end{align}
where 
the expansion parameter $Q$ is related to the natural energy scale of the process and $\Lambda$, the breakdown scale of $\chi$EFT, as $Q={\rm max}\{m_\pi,p\}/\Lambda$, with $m_\pi$ being the pion mass and $p$ the average involved low-energy momentum. 
Here, we choose  $\Lambda\sim 500$~MeV,  $Q=m_\pi/\Lambda$ for $^3$He, and 
$Q=p/\Lambda$ for $^4$He, where we take 
$p\sim 180$~MeV as the average nucleon momentum in the nucleus~\cite{SF_Asia}. 
The expansion coefficients $c_n$ are obtained by calculating the observable at each  fixed chiral order.  
For ${\mathcal O}^{\text{Ref}}$, we take results obtained with the phenomenological AV18+UIX potential.
We obtain similar results if we instead use the leading-order (LO) chiral result as ${\mathcal O}^{\text{Ref}}$.

If the $\chi$EFT expansion is truncated at a given order $k$, the associated uncertainty will be dominated by the first omitted term and can be estimated as
$\Delta_k \sim c_{k+1} Q^{k+1}$.
We assume that the all of the expansion coefficients, $c_n$, are of natural size set by the same scale parameter $\bar c$.
These assumptions are encoded in the use of Jeffrey's prior \cite{Jef39} for $\bar c$ and Gaussian priors for each $c_n$.
Explicitly, the prior probability density functions are
\begin{eqnarray}
\text{pr}(\bar c) &=& \frac{1}{\ln (\bar{c}_>/\bar{c}_<) \bar c} \theta(\bar c - \bar{c}_<) \theta(\bar{c}_> - \bar c) \ , \label{EQ: 13} \nonumber \\    
\text{pr}(c_n|\bar c) &=&\frac{1}{\sqrt{2\pi \bar c}} \ \exp\Biggl[ -\frac{c_n^2}{2\bar c^2} \Biggr] \ , 
\end{eqnarray}
where $\bar c_>=10^{3}$, $\bar c_<=10^{-3}$ and the two theta functions constrain $\bar c_< < \bar c < \bar c_>$, so that the probability distribution $\text{pr}(\bar c)$ is normalizable.
The posterior pdf of $\Delta_k$ can then be obtained as \cite{Fur15, LiMuli22}
\begin{eqnarray}
&&\text{pr}\Bigr (\Delta_k \ \bigr | \ \mathbf{c} \ , \ Q \Bigr ) =\\
\nonumber
&&\frac{ \int \! d\bar{c} \ \text{pr} \Bigl( c_{k+1}=\frac{\Delta_k}{Q^{k+1}} \ \bigr| \ \bar{c} \ \Bigr) \!\! \Biggl[ \! \prod\limits_{n=0}^{k} \text{pr}\Bigl( c_n \ \bigr| \ \bar{c} \ \Bigr) \Biggr] \ \text{pr}(\bar{c})}{ Q^{k+1} \int d\bar{c} \ \Biggl[ \ \prod\limits_{n=0}^{k} \text{pr}\Bigl( c_n \ \bigr| \ \bar{c} \ \Bigr) \Biggr] \! \text{pr}(\bar{c})} \ ,
\label{EQ: 16}
\end{eqnarray}
where the coefficients $\mathbf c = \{c_0,..,c_k\}$ are known from our order-by-order calculation of the observable in $\chi$EFT. 


\textit{Results}---
\begin{figure*}[htb]
\includegraphics[width=0.7\textwidth]{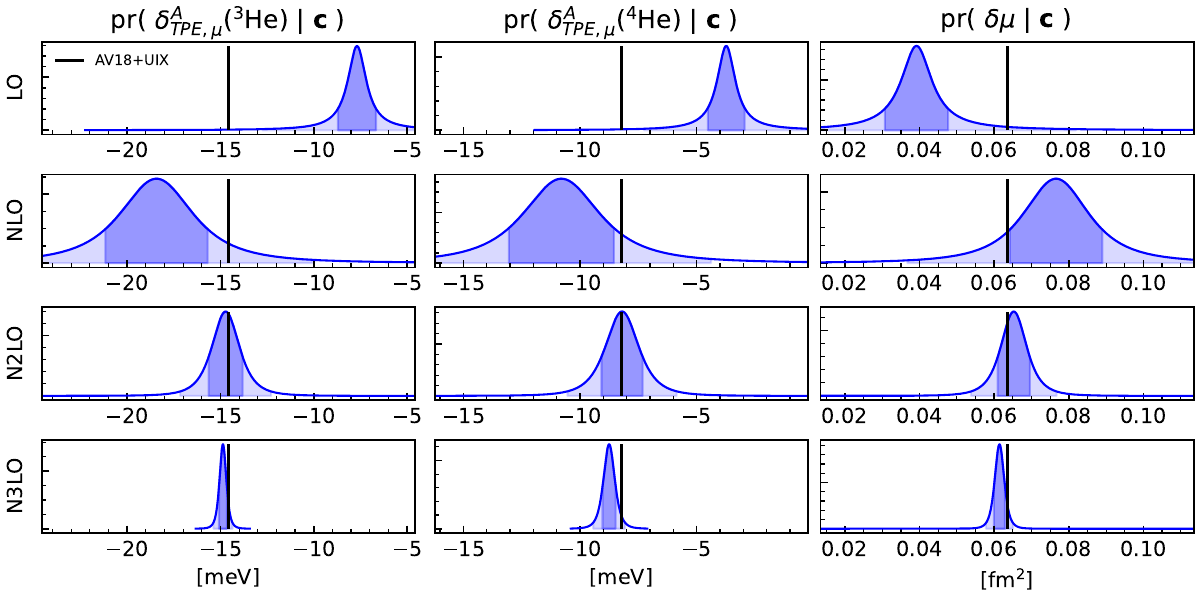}
\caption{Bayesian analysis of the $\chi$EFT expansion in  muonic helium (vertical lines are the  AV18+UIX results~\cite{Ji_18}).}
\label{FIG: 1}
\end{figure*}
In Fig.~\ref{FIG: 1} we show pdfs, obtained via a Bayesian analysis, for values of $\delta^{A}_{\text{TPE}, \, \mu}$ representing our best estimates for the truncation uncertainties of the $\chi$EFT expansion.
The first row shows the uncertainty from truncating the $\chi$EFT expansion at LO, the second row at next to leading order (NLO), the third row at next to next to leading order (N2LO), and the last row at next to next to next to leading order (N3LO).
The dark shaded areas are $68\%$ confidence intervals while the light shaded areas are $95\%$ confidence intervals. 
The vertical black lines are the predictions for $\delta^{A}_{\text{TPE}, \, \mu}$ obtained with the AV18+UIX interaction. 
We note that our LO results are very different from the other orders. This is due to the fact that the expansion coefficients $c_0$ is much smaller than the other coefficients.

In Refs.~\cite{Ji_13,Ji_18}, the uncertainty quantification was performed by comparing the prediction of the N3LO chiral interaction to that of the phenomenological AV18+UIX.
Here, the Bayesian analysis leads to an improved error estimate established on solid statistical ground.
The most precise values are obtained at N3LO, where we find $\delta^{A}_{\text{TPE}, \, \mu}=-14.868(357)$~meV in $\mu^3$He$^+$ and $\delta^{A}_{\text{TPE}, \, \mu}=-8.751(303)$~meV in $\mu^4$He$^+$. The uncertainties are obtained as  quadrature sums of the Bayesians $\chi$EFT expansion error and the remaining sources of uncertainty (e.g.,  $\eta$ expansion,  numerical precision, etc.) quoted in  Refs.~\cite{Ji_18,LiMuli_PhDThesis}. 
Our results can be compared to the previous calculations, $\delta^{A}_{\text{TPE}, \, \mu} =-14.72(31)$~meV in $\mu^3$He$^+$ \cite{Nevo_16,Ji_18} and $\delta^{A}_{\text{TPE}, \, \mu}=-8.49(39)$~meV in $\mu^4$He$^+$ \cite{Ji_13,Ji_18}, from which it is evident that our uncertainty estimation based on Bayesian inference increases the uncertainty in $\delta^{A}_{\text{TPE}, \, \mu}$ of $\mu^3$He$^+$, but reduces the uncertainty in $\mu^4$He$^+$ by 22$\%$. The increased uncertainty in $\mu^3$He$^+$ is mostly due to an update of the uncertainty associated with the $\eta$ expansion~\cite{LiMuli22}.
These results for $\delta^{A}_{\text{TPE}, \, \mu}$ may be combined with remaining terms in Eq.~\eqref{EQ:ns} to produce $\delta_{\rm NS, \, \mu}$.
The charge radii of $^3$He and $^4$He can be found by solving Eq.~(\ref{EQ: 1}) for $r_\text{ch}$  using the measured Lamb-shift~\cite{Kra21,Sch23}.
A summary of these results can be found in Table~\ref{TAB: 1}.
For $^4$He, this constitutes the most precise extraction of the nuclear charge radius up-to-date.

\begin{figure*}[htb]
\centering
\includegraphics[width=0.7 \linewidth]{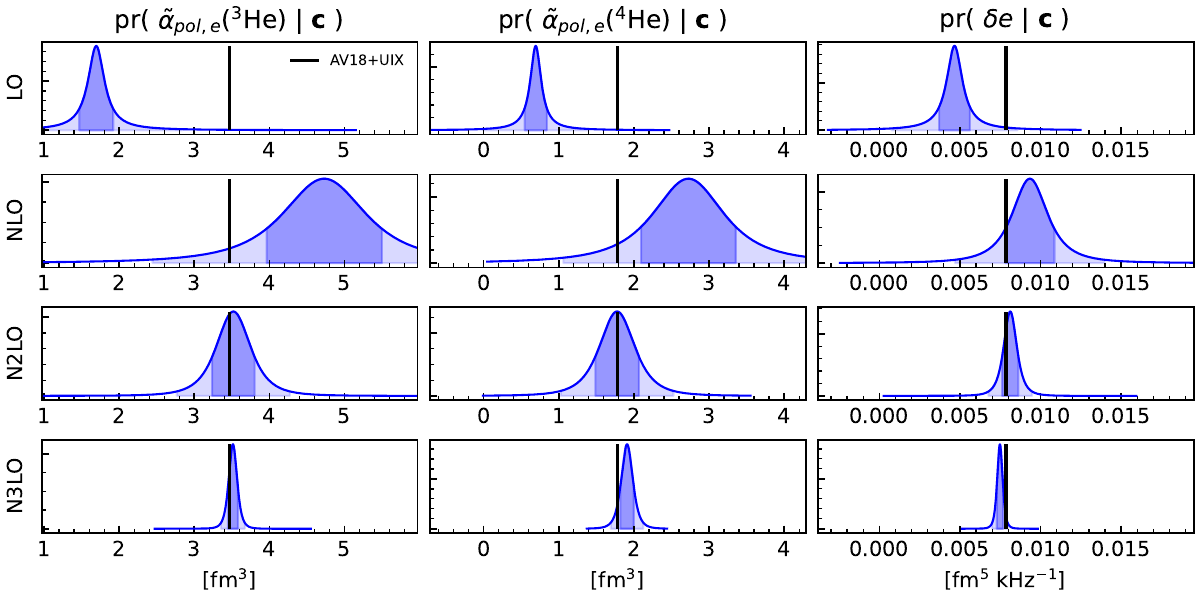}
\caption{Same as Fig.~\ref{FIG: 1} for $\tilde{\alpha}_{\text{pol}}$ in helium atoms. }
\label{FIG: 2}
\end{figure*}

Next, we turn our attention to nuclear structure corrections in ordinary helium atoms, which are calculated here for the first time at various orders in $\chi$EFT.
In Fig.~\ref{FIG: 2}, we show the pdfs for $\tilde \alpha_{\text{pol}}$ representing the evolution of this observable, and of the associated uncertainty, as we increase the order of the $\chi$EFT expansion of the nuclear Hamiltonian. The expansion shows identical features to those of Fig.~\ref{FIG: 1}. At N3LO we obtain $\tilde \alpha_{\text{pol}}=$~3.514(65) and 1.909(78) fm$^3$ for $^3$He and $^4$He, respectively.
Within uncertainties, our results are compatible with the values of 3.56(36) fm$^3$ and 2.07(20) fm$^3$ estimated in Ref.~\cite{Pac07} for $^3$He and $^4$He, respectively. 

\begin{table}[h!t]
\centering
\ra{2}
\begin{tabular}{@{}cccc@{}}
\toprule
& $\delta^{A}_{\text{TPE}, \, \mu}$ [meV] & $\delta_{\text{NS}, \, \mu}$ [meV]  & $r_{\text{ch}}$ [fm] \\
\hline
${}^3$He & -14.868(357) & -15.644(421)  & 1.9704(10)  \\
${}^4$He & -8.751(303) & -9.541(351)  & 1.6793(10) \\
\hline
\end{tabular}
\caption{Nuclear structure effects in muonic helium at N3LO and extracted charge radii.
}
\label{TAB: 1}
\end{table}

\textit{Revisiting the helium isotope shift}-- Starting from our new calculations of nuclear structure effects in atomic spectra, we  update the extraction of $\delta r^2$~\cite{Sch23, Van23}, considering the Lamb shift 2S $\rightarrow$ 2P transition for muonic atoms and the   $2^3{\rm S}\rightarrow 2 ^1 {\rm S}$  isotope shift for ordinary atoms.
First, we invert Eq.~(\ref{EQ: 1}) to obtain the squared charge radius of the individual nuclei, and then we take the difference between $^4$He and $^3$He
    \begin{eqnarray}
    \label{mu_diff}
        \delta r^2 &=& \left[\frac{{\Delta \rm E}(^3\text{He})}{\mathcal{C}(^3{\rm He})} -\frac{{\Delta \rm E}(^4{\rm He})}{\mathcal{C}(^4\text{He})} \right]\label{EQ: diff}\\
        \nonumber 
        &+&\left[\frac{\delta_{\rm QED}(^4\text{He})}{\mathcal{C}(^4\text{He})}
    -        
        \frac{\delta_{\rm QED}(^3\text{He})}{\mathcal{C}(^3\text{He})}\right]\\
        \nonumber 
        &+&\left[
        \frac{\delta_{\rm NS}(^4\text{He})}{\mathcal{C}(^4\text{He})}
         -
        \frac{\delta_{\rm NS}(^3\text{He})}{\mathcal{C}(^3\text{He})}\right] \,.
    \end{eqnarray}
The last term in Eq.~(\ref{mu_diff}) contains the difference of $\delta_{\rm NS}$ for $^3$He and $^4$He.
For $\delta^N_{\rm TPE}$ in Eq.~(\ref{Eq:tpe_sep}) and $\delta_{\text{3PE}}$, $\delta_{\text{EVP}}$, and $\delta_{\text{MSEVP}}$
in  Eq.~(\ref{EQ:ns}), we use the values from Ref.~\cite{Pac23} and analyze the part which we newly calculated, namely
\begin{eqnarray}
  \delta \mu&=&\left[\frac{\delta^A_{\rm TPE, \mu}(^4\text{He})}{\mathcal{C}(^4\text{He})} -\frac{\delta^A_{\rm TPE, \mu}(^3\text{He})}{\mathcal{C}(^3\text{He})}\right]\,,\label{dmu}\\
  \delta e &=&\left[{\tilde{\alpha}_{\text{pol,e}}(^4\text{He})} -
  {\tilde{\alpha}_{\text{pol,e}}(^3\text{He})}\right]  \frac{1}{\mathcal{C}}\,,
\end{eqnarray}
for muonic atoms and ordinary atoms, respectively\footnote{
Note that the constant ${\mathcal C}$ is the same for $^{3,4}$He~\cite{Pac15} for the isotope shift of this transition in ordinary atoms.}. We perform a Bayesian analysis of this difference, which will naturally take correlations between $^3$He and $^4$He into account. 

Results of the statistical analysis of $\delta \mu$ and $\delta e$ for each chiral order are shown in Fig.~\ref{FIG: 1} and Fig~\ref{FIG: 2}, respectively.
At N3LO we obtain $\delta \mu= 0.0614(15)(21)$ fm$^2$, where the first uncertainty comes from the Bayesian analysis of the $\chi$EFT expansion while the second includes all the rest. 
To estimate the latter, from our Bayesian analysis  we extracted the correlation coefficient between the two terms in Eq.~\eqref{dmu}, amounting to 0.8, and used it to propagate the uncertainties to $\delta \mu$. 
Compared to $\delta \mu = 0.0624(42)$ fm$^2$ obtained in Ref.~\cite{Ji_18}, the new result constitutes a significant reduction of the uncertainties due to the inclusion of correlations between  $^{3}$He and $^{4}$He. 
Using values for  $\Delta E$,  $\delta_{\rm QED}$ and  ${\mathcal C}$ reported in Table~\ref{TAB: 2}, and assuming that $\delta_{\text{3PE}}$, $\delta_{\text{EVP}}$, and $\delta_{\text{MSEVP}}$ in $^{3}$He and $^{4}$He are correlated in the same way as $\delta^A_{\rm TPE}$, we obtain 
\begin{equation}
\delta r^2 = 1.0626(29)\text{ fm}^2
\end{equation} for muonic-atoms. 
This improves the uncertainty of the previous determination, $\delta r^2 = 1.0636(31)\text{ fm}^2$  \cite{Sch23}, by about $6 \%$.

\begin{figure}[t]
\includegraphics[width=1 \linewidth]{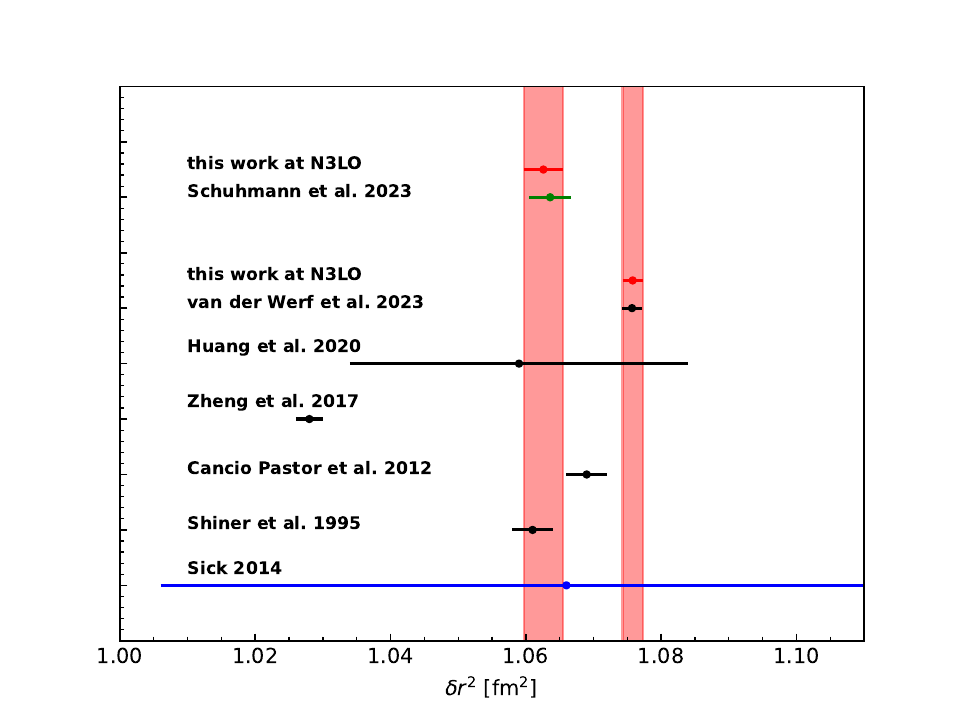}
\caption{Our $\delta r^2$ results at N3LO (red) compared to the 
 muonic helium extraction by Schumann et al.~\cite{Sch23} (green) and the ordinary atom extraction (black) by van der Werf et al.~\cite{Van23}. Shown are also previous ordinary atom extractions (black) by Shiner et al.~\cite{Shi95}, Cancio Pastor et al.~\cite{Can12}, Zheng et al.~\cite{zhe17}, Huang et al.~\cite{Hua20}, and the electron scattering data (blue) by Sick~\cite{Sic14}.
}
\label{FIG: 3}
\end{figure}

For ordinary atoms, we obtain $ \delta e = 0.00748(20)\text{ fm}^5 \text{kHz}^{-1}$ at N3LO, which can be compared with  $ \delta e = 0.0069(19) \text{ fm}^5 \text{kHz}^{-1}$ from Ref.~\cite{Pac07}.
The uncertainty in Ref.~\cite{Pac07} is attributed to higher-order terms in Eq.~\eqref{tpe_e} and Eq.~\eqref{tilde_pol} that were neglected.
This uncertainty was estimated to be of the order of $10\%$~\cite{Pac07},
however there was no attempt to estimate the uncertainties associated with the model dependence of the dipole polarizabity.
In this work, we rigorously quantify the latter, while we assume $10\%$ uncertainty to account for the former, resulting in $ \delta e = 0.00748(95) \text{fm}^5 \text{kHz}^{-1}$.
Using values for  $\Delta E$,  $\delta_{\rm QED}$ and  ${\mathcal C}$ reported in Table~\ref{TAB: 2},  at N3LO we find 
\begin{equation}
\delta r^2 = 1.0758(15)~ {\rm fm}^2.
\end{equation}
We only find a weak modification of the central value compared to the result in Ref.~\cite{Van23}, $\delta r^2=1.0757(15)$, which used Ref.~\cite{Pac07,pat17}. 
This highlights the weak sensitivity of ordinary helium atoms to nuclear polarizabilities.
\begin{table}[h!t]
\centering 
\ra{2}
\begin{tabular}{@{}ccccc@{}}
\toprule
{2S-2P} & $\Delta E$ [meV] & $\delta_{\text{QED}} $ [meV] & $\mathcal{C}$  [meV fm$^{-2}$] \\ 
\hline
${\mu}^3$He$^+$ & -1258.598(48)\cite{Sch23} & -1644.348(8)\cite{Pac23} & 103.383\cite{Pac23} \\
${\mu}^4$He$^+$ & -1378.521(48)\cite{Kra21} & -1668.491(7)\cite{Pac23} & 106.209\cite{Pac23} \\
\hline\hline
{2$^3$S-2$^1$S} & $\Delta E$ [kHz] & $\delta_{\text{QED}}$ [kHz] & $\mathcal{C}$ [kHz fm$^{-2}$] \\ 
{$^3$He-\!$^4$He} & \footnotesize{-5\,787\,729.76(26)\cite{Van23}}  & \footnotesize{-5\,787\,499.05(19)\cite{pat17}}& \footnotesize{-214.66(2)\cite{Pac15}}\\
\hline
\end{tabular}
\caption{Values used in Eq.~\eqref{mu_diff} in the relevant units.  Note that the hyperfine splitting correction to the $2^3$S state in $^3$He is included in the quoted $\delta_{\text{QED}}$ value.
}
\label{TAB: 2}
\end{table}
In Fig.~\ref{FIG: 3}, we present our values of $\delta r^2$  (red) at N3LO in comparison to previous extractions by Schumann et al.~\cite{Sch23} and van der Werf et al.~\cite{Van23}, along with other  experimental results, also shown in Fig.~5 of Ref.~\cite{Sch23}. 
While the van der Werf et al.~datum remains mostly unchanged due to the insignificance of nuclear structure corrections in  ordinary atoms, our updated analysis moves the muonic atom datum the the left, enhancing the discrepancy to a $4~\sigma$ level.


\textit{Conclusions}---
We have calculated nuclear structure corrections in muonic and ordinary helium atoms using accurate few-body methods and $\chi$EFT interactions at various orders.
With respect to muonic atoms, while various chiral orders were previously explored for muonic deuterium~\cite{Hernandez_14,Hernandez_2018},  this is accomplished here  for three- and four-body nuclei. 
Ordinary helium atoms are  analyzed here from a consistent theoretical point of view for the first time. 

We applied Bayesian inference techniques to quantify uncertainties stemming from the  $\chi$EFT truncation. 
For muonic atoms, this allows to improve previous simple estimates~\cite{Ji_18} based on comparing the N3LO chiral  and the AV18+UIX phenomenological interactions. 
At  N3LO, the obtained Bayesian  uncertainty is comparable to the previous ones, but is now founded on solid statistical ground. 

To bring the muonic-atom extraction of  $\delta r^2$  in agreement with that from ordinary atoms, $\delta \mu$ would have to change by
$0.00875$ fm$^2$. This change is excluded by our Bayesian analysis at the $95 \%$ confidence level. 
Conversely, to bring ordinary atom measurements in agreement with the results from muonic atoms, $\delta e$ would have to be 8 times larger as well as the opposite sign. 
This is excluded by our Bayesian analysis, as well as from the evidence that $^4$He is more strongly bound  than $^3$He and therefore a change of sign is not expected.

Overall, the helium isotope shift puzzle, see Fig.~\ref{FIG: 3}, is not resolved, but rather enhanced to a 4$\sigma$ level. 
Therefore, our theoretical analysis suggests that most likely the explanation of the puzzle is related to underestimated systematic errors in the experiment or neglected correction terms.
Finally, the charge radius  extracted for  $^4$He with our new nuclear structure corrections (see Table~\ref{TAB: 1}),  constitutes the most precise value to date.


\vspace{0.3cm}
\begin{acknowledgements}
We acknowledge useful discussions with N.~Barnea, R.~Pohl, and K.~Pachucki.
This work was supported  by the Deutsche
Forschungsgemeinschaft (DFG)
through the Cluster of Excellence ``Precision Physics, Fundamental
Interactions, and Structure of Matter" (PRISMA$^+$ EXC 2118/1) funded by the
DFG within the German Excellence Strategy (Project ID 390831469).
\end{acknowledgements}


\bibliography{mybibfile}

\end{document}